\documentclass[12pt]{iopart}
\usepackage{iopams}  
\usepackage{natbib}
\usepackage{color}
\usepackage{ulem}
\usepackage{graphicx}
\usepackage{aas_macros}

\newcommand{\DHA}[1]{\textcolor{black}{#1}}

\begin{document}
\title[Ambient radioactivity induced background in W-TES]{
Radioactivity induced dark count rate for single near-infrared photon detection with a tungsten transition edge sensor at 80 mK}
\author{N. Bastidon\footnote{Present address: Department of 
Physics and Astronomy, Northwestern University, Evanston, IL, USA}, D. Horns}
\address{Institut f\"ur Experimentalphysik, Universit\"at Hamburg, Hamburg, Germany}
\ead{dieter.horns@physik.uni-hamburg.de}
\vspace{10pt}
\begin{indented}
\item[]February 19th 2017
\end{indented}
\begin{abstract}
The intrinsic background count rate of tungsten superconducting 
 transition-edge sensors (TES) is low, and the calorimeters
using these sensors can resolve the energy of single photons. These facts
make the sensors particularly interesting for the background-limited
searches of new processes
and particles. In this contribution,
the intrinsic background of a tungsten  TES has been
investigated. After excluding other sources (e.g., cosmic muons, 
thermal background) to be relevant for the observed background rate of
$10^{-4}$~s$^{-1}$ for the detection of photons with
a wave length of 1064~nm, we investigate the impact of
natural radioactivity.  Dedicated 
measurements using gamma-emitters mounted outside the cryostat have been used
to estimate the sensitivity of the TES setup for ionizing radiation. We have
found that indeed an increased background can be observed in the presence of the
radioactive sources. After selecting events which populate our signal region tuned for single photon detection at near-infrared, roughly 0.5~\% of the 
events produced by gamma-rays appear indistinguishable from those due to single photons with
1064~nm wave length. This ratio is consistent with  that
 observed for the residual background
detected with the TES at a rate of $10^{-4}$~s$^{-1}$. 
From this, we conclude that the bulk of the observed 
background count-rate in the signal region can be explained by natural radioactivity. 
\end{abstract}
\vspace{2pc}
\maketitle
\ioptwocol
\section{Introduction}
The development of single-photon and photon-number resolving detectors has been largely driven by the field of quantum information science and metrology (see Fig. 1 in \citet{doi:10.1063/1.3610677}). 
{While in these applications, the requirements of e.g., high count-rate and high efficiency are 
most important, searches for rare processes benefit from minimized instrumental background.}
{Recently,
an experiment has been proposed to search for new light fundamental bosons \cite{2013JInst...8.9001B} (see the same reference for an
in-depth discussion of the
underlying physics of axion-photon mixing and the experimental setup).
 A crucial component of the experiment is a low-background 
(required dark count rate of $10^{-6}$~s$^{-1}$) and highly efficient 
(required overall detection 
efficiency of 75~\%)  single-photon detector sensitive at a wave length of $1064$~nm.}
\DHA{While the required detection efficiency has been
demonstrated in previous studies \citep{2008OExpr..16.3032L}, the dark 
count rate of a TES exceeds the requirement:  In a
previous study of the same sensor as used here (a 25~$\mu$m square
and 20~nm thick tungsten film) a dark count rate of $10^{-4}$~s$^{-1}$
had been found \citep{2015JMOp...62.1132D}.}
\\
\DHA{Here, we investigate the intrinsic dark count rate of a TES in order
to clarify the origin of the measured dark count rate. The
paper is structured as follows: in the next section, we describe the
experimental setup and list the potential origin of dark counts. In 
Section 3, we give a summary of the measurements carried out followed
by Section 4, where the results are interpreted. We conclude on 
potential approaches to reduce the background further.}

\section{\DHA{Experimental setup}}
 
\begin{figure}
 \includegraphics[width=0.56\linewidth]{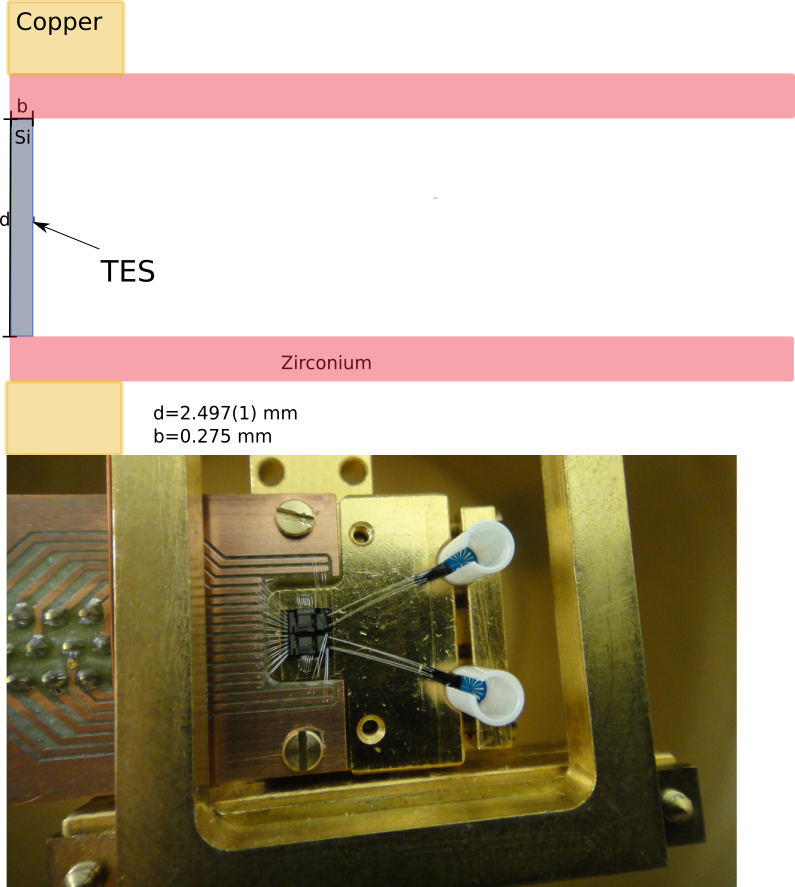}
 \includegraphics[width=0.40\linewidth]{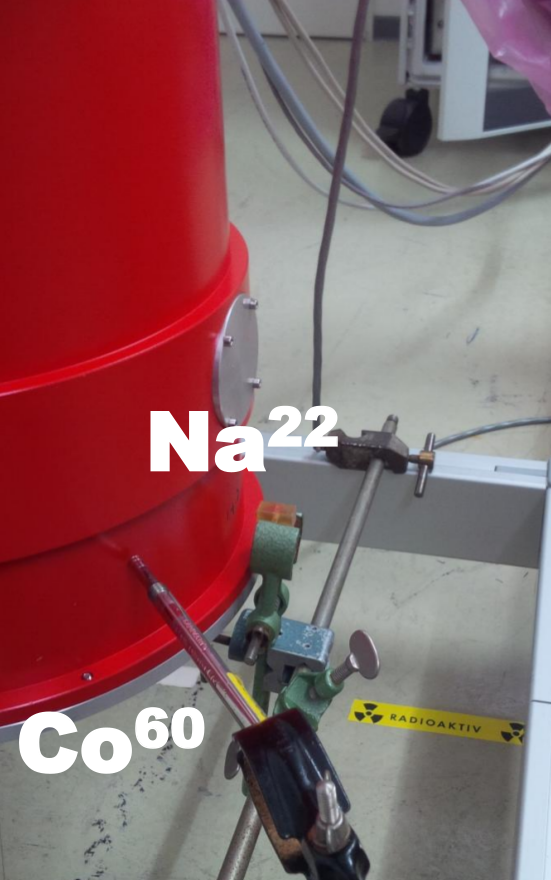}
\caption{\label{Fig1} Left panel: A photography of the TES detector assembly and a sketch 
of the relevant components.
Right panel: The two radioactice sources mounted outside the outer
aluminum shielding of the ADR.}
\end{figure}
 The structure of the TES consists of a 20 nm thick layer of W 
sandwiched in a dielectric stack consisting of 2~nm amorphous
Si and 127/132~nm layers of SiNx on top of a 80~nm thick Ag coating.
The stack is optimized
to reduce the relative reflectance to 0.32~\% at $\lambda=$1064 nm. The reflectance slowly rises
to larger values for  \DHA{$\lambda<800$} nm and  \DHA{$\lambda>1500$~nm.} The active area 
is \DHA{a} square  with 25 $\mu$m side length. \DHA{It}   
resides
on a silicon substrate of 275~$\mu$m thickness and
diameter of 2.5~mm (see Fig.~\ref{Fig1}, \DHA{left panel for a photography and a sketch of the assembly mounted on the 
cold finger}). \DHA{The silicon substrate fits inside a zirconium sleeve which 
provides a defined connection to a single mode fiber held in a zirconium
ferrule. However, in the setup considered here, the sleave is empty.}\\
 \DHA{The TES is stabilized at its transition temperature of 80~mK via negative
thermo-electric feed-back. 
The TES is read-out via a two-stage squid optimized to 
operate in typical ambient magnetic field of $10^{-4}$~T \citep{2007ITAS...17..699D}}.  The
resulting signals are digitized and stored using a digital oscilloscope.\\
The TES is mounted inside a cryostat cooled by a two-stage pulse-tube cooler to
reach 3 K followed by a two-stage
adiabatic demagnetization system. 
\\
In order to understand the origin of the indistinguishable 
background events registered in the signal region 
defined for 1064 nm photons, we have carried out 
dedicated measurements with radioactive sources mounted outside
the cryostat at a distance of $d_{source}=(0.2\pm0.02)$~m to the detector \DHA{(see
Fig.~\ref{Fig1}, right panel for a photography of the setup)}. 
We used two sources ($^{60}$Co with 
an activity of $A_\mathrm{Co}=53$~kBq
and $^{22}$Na with an activity of $A_\mathrm{Na}=280$~kBq). 
The sodium source produces a 
gamma-ray spectrum with a 511~keV annihilation
line from the $\beta^+$ decay of the $^{22}$Na into $^{22}$Ne$^*$ 
which emits a 1.275 MeV photon to relax into the ground state. 
The  \DHA{$^{60}$Co}   \DHA{nuclei} decay via $\beta^-$ to Ni, populating
two excited states which lead to the emission of two photons with
1.17~MeV and 1.3325~MeV. We have simulated the setup with GEANT4 \citep{2016NIMPA.835..186A}
and found on average approximately 2 additional Compton scattered electrons
with energies above 100 keV that are released inside the aluminum shielding 
for every gamma-photon impinging
from the outside. We have verified  this simulation result using a 
NaI(Tl) based gamma-ray spectrometer mounted inside
the aluminum shield while the radioactive source was mounted 
outside of the shield.  \\
We accumulated events with either source individually and 
with the combination of both sources. The sources faced the detector
inside the cryostat. We also took background data without a source. \DHA{In a different setup, we have determined the response of the TES to an attenuated flux of photons from a 
laser coupled to a single-mode fiber which guides the light to the TES surface.}
\section{Data analysis}
\begin{figure*}
 \includegraphics[width=0.5\linewidth]{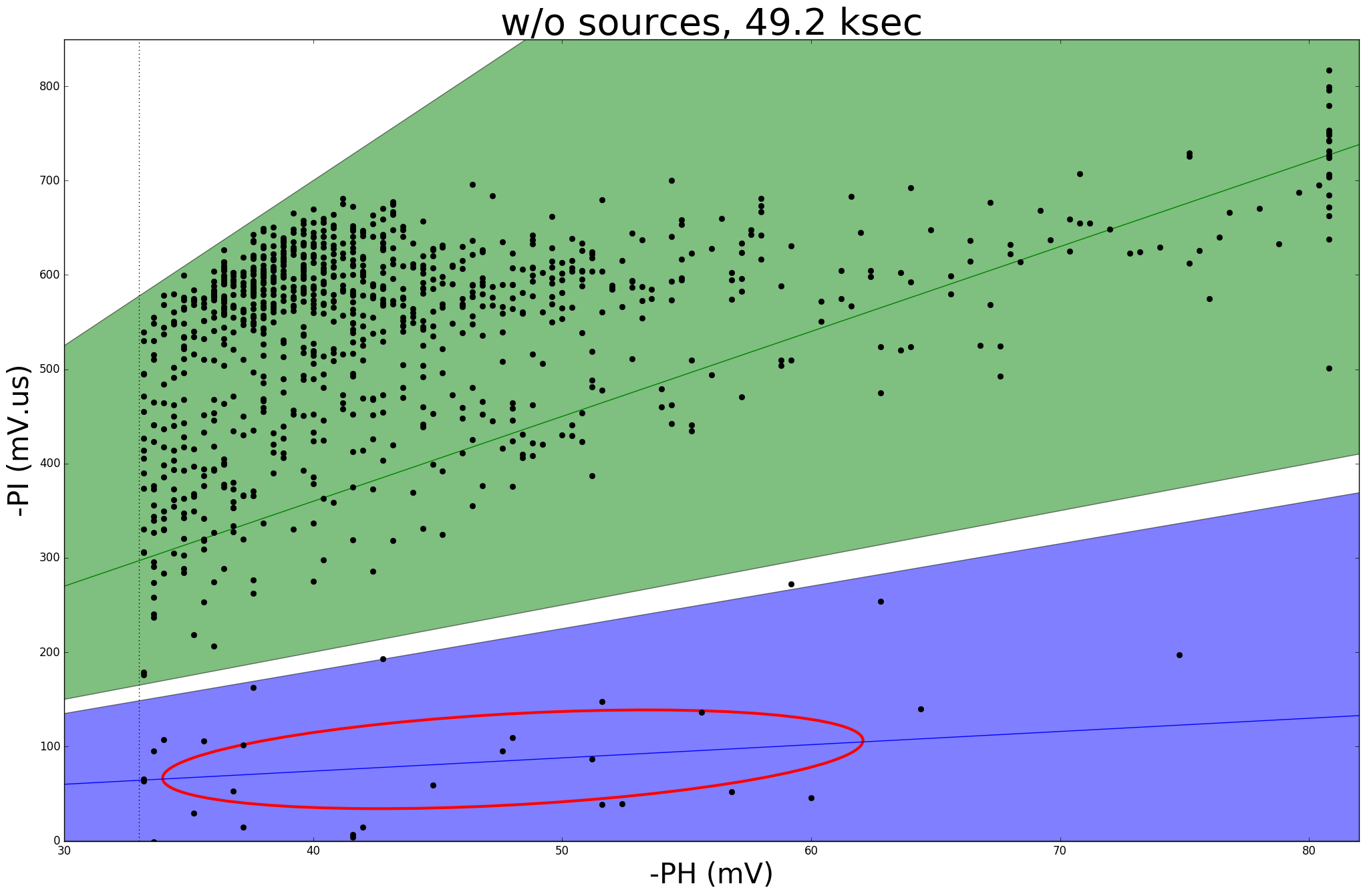}
 \includegraphics[width=0.5\linewidth]{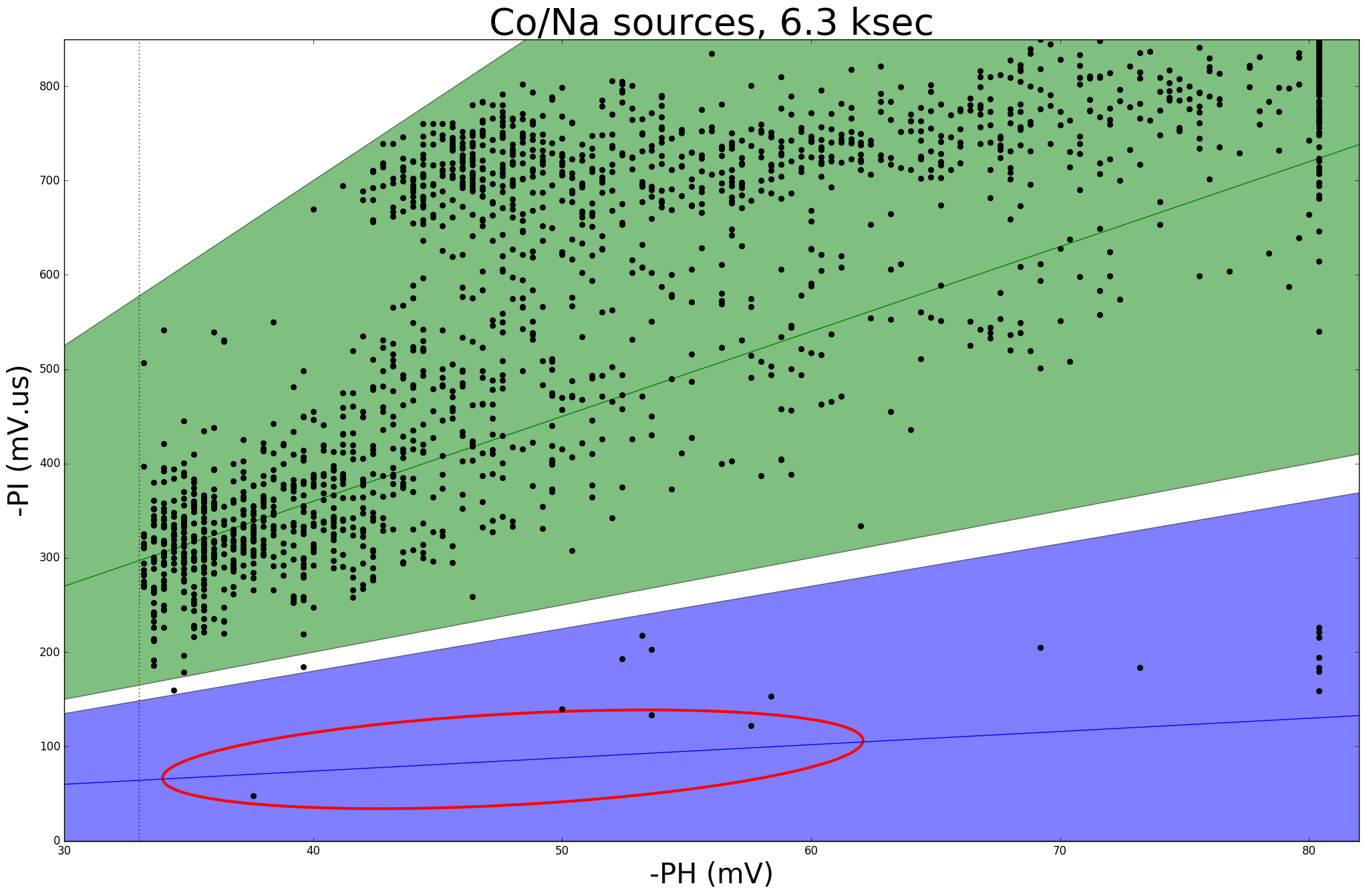}
\caption{The left and right panels display the measured distribution of pulse
height (PH) and pulse integral (PI) for a long duration measurement without
\label{back}
radioactive source (left panel) and a shorter measurement with two radioactive sources mounted outside the ADR. The green and blue shaded regions indicate
coarsely the photon like (green, lower band) and background like (blue, 
upper band) regions. The ellipse marks the region in which photons with
a wavelength of 1064~nm are selected. }
\end{figure*}

\DHA{For each event passing a trigger threshold of 32~mV,  the time-line of the output voltage has been recorded in 20~ns steps to cover a total of 20~$\mu$s. The data are analysed offline to 
determine the peak voltage (pulse height amplitude: PHA) and the pulse integrale (PI). For calibration purposes, we determine the locus of the events 
and an ellipse encompassing the signal events in a plane of 
PI and PHA from a run 
 where a photon source at $\lambda=1064$~nm was coupled to a single mode fiber which guides the photons to the TES. The ellipse is chosen 
to encompass 97.2~\% (3~$\sigma$) of the laser induced events. We have verified that the locus of the event distribution
does not change when maintaining similar settings of the working-point of the TES and 
fiber positioning, for further details on this calibration step see 
\cite{dreylingeschweiler,2015JMOp...62.1132D}.\\
In order to determine the dark count rate of the TES, we have disconnected the fiber and
operated the TES in a similar mode as previously. After an exposure of $49.2\times10^3$~s, the resulting distribution of background events is shown in Fig.~\ref{back}, left panel. Finally, we show in Fig.~\ref{back}, right panel, 
the events registered with
the sodium and cobalt sources mounted outside the ADR.  The distribution 
of events falls in two regions marked in the PI-PH planes. The green band
encompasses short pulses with an effective pulsewidth ($PI/PH\approx 2~\mu$s),
while a larger number of events is characterized by a longer effective
pulse width ($PI/PH\approx 10~\mu$s). When comparing the two measured distributions, the overall shape is similar. Some differences are noteworthy: in the
case of the MeV-energy of the radioactive gamma-rays of the sodium and cobalt
sources, the long duration events are on average depositing more energy. This
may be the result of different energy losses of lower-energy ambient gamma-rays which tend to loose energy predominantly via photoelectric interaction while
at MeV energies, Compton scattering starts to dominate.} 

\section{Results}
The results of the measurements carried out are summarized in \Tref{result}. The 
numbers listed have been extracted after selecting the pulse height (PH)
and pulse integral (PI) of events  as shown in Fig.~\ref{back}.

\begin{table}
\caption{\label{result}
Summary of the data taken with and without radioactive sources. The errors shown
are statistical only.
}
\begin{indented}
\item[]\begin{tabular}{rrrrr}
\br
  Source & activity $A$ & time          & $\dot n_\mathrm{trigger}$   & $\dot n_\mathrm{signal}$ \\
   name  & [kBq]        & [10$^3$~s]& $10^{-3}$~s$^{-1}$          & $10^{-3}$~s$^{-1}$  \\
\mr
  $^{60}$Co    & 53            & 5.4       & $66(4)$                     & 0.4(3)  \\
  $^{22}$Na    & 280           & 6.3       & $256(6)$                    & 1.3(4)  \\
  $^{60}$Co,$^{22}$Na & 333 & 6.3       & $291(7)$                    & 0.5(3) \\
  \DHA{w/o src.} &unkn.        & 49.2      & $23(7)$                     & 0.14(5) \\
\br
\end{tabular}
\end{indented}
\end{table}

\subsection{Triggered and signal rate increase}
In our case, we measure a \textit{raw rate} (before selecting signal-like
events) increase 
$$\Delta \dot n_\mathrm{trigger}= \dot n_\mathrm{trigger}-
\dot n_\mathrm{trigger}^\mathrm{w/o~ src}$$ 
normalized to the activity of the source of
$$\frac{\Delta \dot n_\mathrm{trigger}(\mathrm{Co})}{A_\mathrm{Co}}
=(0.8 \pm 0.1)\times\frac{10^{-3}~\mathrm{s}^{-1}}{\mathrm{kBq}}$$
$$\frac{\Delta \dot n_\mathrm{trigger}(\mathrm{Na})}{A_\mathrm{Na}}
=(0.81 \pm 0.03)\times \frac{10^{-3}~\mathrm{s}^{-1}}{\mathrm{kBq}}$$
$$\frac{\Delta \dot n_\mathrm{trigger}(\mathrm{Na}+\mathrm{Co})}{A_\mathrm{Na}+A_\mathrm{Co}}
=(0.80\pm 0.03)\times \frac{10^{-3}~\mathrm{s}^{-1}}{\mathrm{kBq}}$$ 
for 
the two different sources and their combination.
The increase of the activity normalized rate is consistent for the two different types of 
radioactive sources.
Note, this is not trivial given that the energy spectra of the emerging photons
are quite different.
\\
The majority of these additional events however are filtered out
by their distinctively different pulse shape.
 After selecting events which
fall into the signal region selected for 1064~nm photons, the
rate remains increased. We find  for 
the excess rate
$$\Delta \dot n_\mathrm{signal}=\dot n_\mathrm{signal}-\dot n_\mathrm{signal}^\mathrm{w/o~src}$$
normalized to the activity $A$: 
$$\frac{\Delta \dot n_\mathrm{signal}(\mathrm{Co})}{A_\mathrm{Co}} =
(5\pm5)\times \frac{10^{-6}~\mathrm{s}^{-1}}{\mathrm{kBq}}$$
$$\frac{\Delta \dot n_\mathrm{signal}(\mathrm{Na})}{A_\mathrm{Na}} =
(4\pm1)\times \frac{10^{-6}~\mathrm{s}^{-1}}{\mathrm{kBq}}$$
$$\frac{\Delta \dot n_\mathrm{signal}(\mathrm{Na}+\mathrm{Co})}{A_\mathrm{Na}+A_\mathrm{Co}}=
	(1.1\pm 0.9)\times \frac{10^{-6}~\mathrm{s}^{-1}}{\mathrm{kBq}}.$$
Clearly, longer exposures are needed to determine these rates more accurately\footnote{Prior to these measurements, the crystal unit of the ADR deteriorated leading to short holding times.}.
The number of signal-like events accumulated in all three measurements is
$n_\mathrm{signal}=13$
 with an expected background of 2.5 events which corresponds
to a significance of $\approx 3$ standard deviations using the likelihood
method given by Eqn. 17~in \citep{1983ApJ...272..317L}. 

Taking this excess to be significant, we can estimate the ratio 
of events which contaminate the signal region to the triggered event number.
We call this quantity $\xi = n_\mathrm{signal}/n_\mathrm{trigger}$.
For the background data sample (without radioactive source), 
we find a ratio of $\xi(\mathrm{background})=(0.6\pm0.3)~\%$. This can be compared
readily with the  \DHA{corresponding ratio 
for the data samples taken with radioactive sources} taking the combination of the 
three data sets $\xi_\mathrm{Na,Co}=(0.3\pm0.1)~\%$, consistent with 
$\xi(\mathrm{background})$ within the uncertainties. 

From this estimate, we can draw the conclusion, that the observed fraction of 
signal-like events in our background sample is consistent with the one observed
for events generated by energetic particles hitting the cryostat.

\subsection{Detection efficiency for energetic photons}

We can use the measurement to determine the efficiency for the detection of
energetic photons impinging from outside the cryostat. Assuming for simplicity that 
we can estimate the $\gamma$-ray flux $\phi_\gamma$ at the location of the sensor 
at distance $d$ related to the activity of the source $A$ to
be 
\begin{eqnarray}
 \phi_\gamma =  \frac{A}{4\pi d_\mathrm{source}^2}.
\end{eqnarray}
We can then proceed to estimate the effective area 
$\pi r_\mathrm{eff}^2$ of the TES for detecting these gamma-rays 
by requiring
\begin{eqnarray}
 \frac{\Delta \dot n}{A} = \frac{r_\mathrm{eff}^2}{4 d_\mathrm{source}^2},
\end{eqnarray}
and therefore calculating an effective radius $r_\mathrm{eff}=2d\sqrt{\Delta \dot n/A}=(350\pm36)~\mu$m.
Given the side length of the active sensor of $20~\mu$m, this implies that most gamma-rays must 
be detected via the coupling of the sensor to the substrate  where most of the energy of the gamma-rays will
be deposited. Going one step further, the effective radius for signal-like events is reduced by a
factor of $\sqrt{\xi}\approx 10^{-1}$ or correspondingly a radius  similar to the
geometrical extension of the sensor.

\section{Discussion}
The obvious result of this measurement is that we detect clearly the presence
of the radioactive sources mounted outside the cryostat 
as an addition to the background rate. Through this measurement, we can quantify
the fraction of ionising events which are registered with the TES and appear identical
to a $\lambda=1064$~nm single photon. This fraction ($\xi=n_\mathrm{signal}/n_\mathrm{trigger}$),
is very similar to the one determined from observations without radioactive sources mounted. 
This would be consistent with the hypothesis that the insuppressable background rate of $10^{-4}~\mathrm{s}^{-1}$ is caused by ionizing radiation from ambient radioactivity.\\ The
efficiency for detecting gamma-rays is quantified by an effective radius
$r_\mathrm{eff}$ up to which photons and Compton scattered electrons 
are detected when impinging on the
substrate. The estimated $r_\mathrm{eff}=(350\pm36)~\mu$m is much larger than
the geometrical extension of the TES which implies that energy deposited in the
Si substrate is indirectly detected with the TES 
very likely through heating. A very small fraction of
these events is finally detected with signal properties identical to photons
 with a
wavelength of 1064~nm. 
The similarity of these events implies a
very similar type of energy deposition and heating/cooling of the TES - 
even though the energy is initially released as ionization. \\
The measurement carried out  indicates that the signal-like events are very likely the result of 
secondary effects related to the energy deposition and possibly
even re-radiation of locally heated substrate. \\
Finally,  we can estimate the background rate using the detection efficiency
determined from our measurements. We assume the level of ambient radioactivity
to be similar as determined in \citep{2014NIMPA.745....7B} 
which translates into an isotropic flux of 
$I_\gamma=2\times 10^{4}$ photons~m$^{-2}$s$^{-1}$~$(4\pi)^{-1}$ 
for energies larger than 100~keV. 
Given that the value found for $r_\mathrm{eff}\approx 275~\mu\mathrm{m}$ (roughly the thickness 
of the substrate), we assume that the setup has a
sensitivity close to isotropy and therefore can readily estimate the background event rate
 $\dot n_\mathrm{trigger, \gamma}= (8\pm2)\times 10^{-3}~\mathrm{s}^{-1}$, 
which accounts for roughly a third of the measured background rate listed in Table~1. 
Note, additional
background will be induced by Compton-scattered electrons from the shield (see also previous section)
as well as from radioactive nuclei present in the ambient material.

\section{Conclusion}
 The measurement described in this article has helped to understand the
intrinsic dark count rate of a transition edge  sensor
operated at 80~mK. A small fraction $\xi \approx 0.5~\% $ of the measured
background events is indistinguishable from individual photons detected at 1064
nm wavelength. Based upon our preliminary results, we
find  that a large fraction of the residual background is
produced from ambient radioactivity - mainly gamma-rays that can deposit energy
in the TES or its environment in such a way that the pulse shape is  indistinguishable from that in single photon events. 
Other sources of background (e.g., ambient light or cosmic muons have been 
excluded \citep{Bastidon2017}). The underlying mechanism for pulse generation is
however not yet identified. It seems plausible, that local heating of the substrate could either
propagate through phonons to the TES layer or more indirectly, via photons emitted from the surface. 
Future studies are required to find a remedy to reduce this dark count rate even further 
in order to improve the sensitivity of TES detectors for rare process searches. In principle, the reduction of background
could be achieved through careful selection of the materials (radio purity) in the immediate vicinity of the detector as well as
by installing additional shielding. For an optimized design of the detector assembly, a deeper investigation of the
mechanism of signal generation is required.

\section*{Acknowledgment}
 We acknowledge the NIST group, especially Sae Woo Nam and Adriana Lita for providing us with 
the TES samples. NB acknowledges the support of the PIER graduate school. DH 
acknowledges the support of the SFB 676 cooperative research center funded
by the DFG. 

\newcommand{\newblock}{}
\bibliographystyle{agsm}
\bibliography{mybib}
\end{document}